\documentclass[epj]{svjour}
\usepackage{amsmath}
\usepackage{amssymb}
\usepackage{graphicx}
\begin{document}

\title{%
Selective advantage of topological disorder in biological evolution
}%
\author{%
Michal Kol\'{a}\v{r}\inst{1,2}\thanks{e-mail: {\tt kolarmi@fzu.cz}
(corresponding author)}
\and
Franti\v{s}ek Slanina\inst{1}\thanks{e-mail: {\tt slanina@fzu.cz}}
}%
\institute{%
        Institute of Physics,
    Academy of Sciences of the Czech Republic,
    Na~Slovance~2, CZ-18221~Praha,
    Czech Republic
\and
    Institute of Physics of Charles University in Prague,
    Ke Karlovu~5,
    CZ-12116~Praha, Czech Republic
}%
\date{%
(This version was processed on: \today )
}
%
\abstract{%
We examine a model of biological evolution of Eigen's quasispecies
in a so-called holey fitness landscape, where the fitness of a site is
either $0$ (lethal site) or a uniform positive constant (viable
site). The evolution dynamics is therefore determined by the topology of
the genome space which is modelled by the random Bethe lattice. We
use the effective medium and single-defect approximations to find
the criteria under which the localized quasispecies cloud is
created. We find that shorter genomes, which are more robust to
random mutations than average, represent a selective advantage
which we call ``topological''. A way of assessing empirically the
relative importance of reproductive success and topological
advantage is suggested.
\PACS{
      {05.40.-a}{Fluctuation phenomena, random processes, noise, and Brownian motion}   \and
      {87.23.Kg}{Dynamics of evolution}  \and
      {72.15.Rn}{Localization effects (Anderson or weak localization)}
     } 
} 
\maketitle

\section{Introduction}\label{introduction}

The mechanism of biological evolution is a very challenging topic
for the physical community. This is well expressed in the numerous
models of biological evolution that have emerged in recent years. The
list of  models studied starts with large-scale properties of  the
evolutionary  process---massive global extinc\-tions---and ends
with the works which are aimed at following the replicative behavior
of the chemical structures holding information, namely DNA molecules. A lot
of effort has been devoted to this area recently
\cite{ba_fly_la_92,ba_sne_93,pa_ma_ba_96,va_au_95,fe_pla_dia_95,so_ma_96,ro_ne_96,zhang_97,drossel_98,sol99,sla_kot_99,sla_kot_00,kot_sla_sta_02,nim_cru_huy_99,wilke_01,wilke_01a}
and still new fruitful ideas emerge
\cite{kra01,chri_coll_hall_jen02,hall_chri_coll_jen02}.

The process of biological evolution consists in three steps:
$reproduction\rightarrow mutation\rightarrow selection.$ The
important thing to note is that the \emph{biological fitness}
function which denotes an individuals' ability to produce viable
offsprings depends on their phenotype. On the other hand, the
mutations occur in the  genotype---information stored in a
sequence of DNA. The fitness function, which assigns to each
microscopic genotype its ability to reproduce and pass on to the
next generation, shows an overwhelming complexity. This property
makes the theoretical treatment of evolution an extremely
complicated task.

Simplifications of the problem are necessary. The fruitful idea of
adaptive landscapes was introduced by Wright \cite{wri31} and was
later further simplified to fitness landscapes later. In this scheme the
individual is represented as a point in a multidimensional space and
to every point a fitness value is assigned. Thus, a
landscape is formed of mountains of genetically adapted positions
and valleys of lethal genomes. Note the idealization: the fitness
is directly given by the genome of the individual, not by its
(extended) phenotype.

In fact, the fitness landscape is not static since it strongly
depends on the ever changing environment, which includes
interactions with (co)evolving species as well as abiotic
influences \cite{kauffman_90a,kau91,wil01}. Evolutionary process
manifests itself the ascent of individuals to peaks on the
fitness landscape.

A broad set of different fitness landscapes was recently used to
study the behavior of the evolutionary system. These models employ
both static \cite{pel01,alt01} and dynamic landscapes
\cite{wil01,nil00,pekalski_00,pek_wer_01}. They include the
\emph{sharply-peaked landscape} (SPL) with a single preferred
genome (wildtype), the \emph{Fujiyama landscape}, or the
\emph{holey landscape} (HL). Several recent reviews summarize
various approaches explored \cite{wil01,baa99,pel97,drossel_01}.

Generally, one simplifies the genetic code considering only a 
two-letter $\{0, 1\}$ alphabet. Let us consider, for example, the digits
$1$ and $0$ as symbols for two different alleles of a certain
gene\footnote{Other possibility is to assign pyrimidines in the
genetic code by $1$ and purines by $0$.}. Assuming constant genome
length of $d$ loci the state space is a hypercube  ${\mathcal
G}=\{0,1\}^{d}$.

The dynamics of evolution on fitness landscapes have the most
prominent scheme in the \emph{quasispecies} model as introduced by
Eigen \cite{eig71}. Originally it was introduced as a model for
chemical prebiotic evolution, but it showed itself plausible in the
 investigation of the mechanisms of microevolution of viruses and
bacteria, i.e. organisms with relatively simple genomes.

The quasispecies is defined as a cloud of closely related
individuals. They hang together but certainly they may move on the
fitness landscape. We assume an infinite population size and
therefore the dynamics of the quasispecies are probabilistic, but
without any noise which would be induced by finite size effects.
The quasispecies obeys three basic processes: reproduction,
selection and is liable to mutational genetic changes.
Reproduction and selection are treated together---the reproductive
ability depends on the fitness and hence selection takes part
here. Mutation can occur in the individual's genome with a
probability rate $\mu$.

It is evident that the native geometry of the genome space is the
above mentioned hypercube ${\mathcal G}=\{0,1\}^d$. So, this is
the natural starting point when building a model for a fitness
landscape. Recently, the main focus has been aimed at the presence or
absence of the adaptive regime induced by a single maximum in the
fitness landscape. Therefore, the first thing to try is the
sharply-peaked landscape on the hypercube. This model was solved
exactly by Galluccio et al. \cite{ga_gra_zha_96,galluccio_97}.

The most important feature found in the quasispecies model in
SPL is the error threshold \cite{tar92,fra97}. It is the phase
transition that separates two regimes of the quasispecies
evolution, namely the adaptive regime and the wandering regime. In the
adaptive regime the localized cloud of quasispecies is formed
around the wildtype (for us a certain site in the lattice),
whereas in the wandering regime no quasispecies is formed due to a
mutation load.

The error threshold phenomenon comes out of the competition
between the selective advantage of a certain wildtype and the
mutation load presented by the rate $\mu$.  The threshold is then
characterized by the specific value of the selective advantage. We
want to show that the selective advantage in the specific site is
not the only parameter, whose value can distinguish two
significantly different regimes. There are also geometrical
properties, e.g. connectivity of the site, that can make the
genotype in the specific site advantageous. These are the main
tasks of this article.

Indeed, the real landscapes are far more complicated than SPL. It
is expected that rugged landscapes with many competing maxima
represent a realistic picture \cite{ba_fly_la_92}. Such landscapes
are well-known in the theory of spin glass\-es \cite{me_pa_vi_87}
and a ``spin-glass'' theory of evolution was investigated, e.g. in
\cite{am_pe_sa_91}. In our previous work we have studied several
similar  models of the fitness landscape, too \cite{kol_sla_02}.

Yet another approach to the modelling of the fitness landscape is
used for computations \cite{gav98,gav99}---the so-called holey
landscape (HL), where the fitness is either a positive constant
(which may be set to $1$) or it is $0$ which means that the
individual with the corresponding genetic code dies with a
probability of $1$ without having offspring.

Indeed, a large part of the point mutations which may occur at the
basic level of the evolutionary picture are lethal for the
individual---they lead to the `lethal' sites. Therefore, the
hypercube does not represent a good approximation to the
evolutionary dynamics, because only a small part of its edges
represent paths to possible new `hospitable' genomes.

A sparsely connected set of points selected at some of the
hypercube corners is perhaps a better choice. Such sparse graphs
and their adjacency matrices are under study extensively at
present; see e.g. \cite{bi_mo_99,monasson_99,sem_cug_02} and
references therein. The authors observe the localization of the
eigenvectors of the sparse matrices due to topological
characteristics only.

We approach the problem from a somewhat different perspective which
resembles the study of topologically disordered solids where the
random network of bonds is often well modelled by the Bethe
lattice \cite{wil_mas_vel_86}. This, of course, supposes that
there are no short loops in the graph. Supposing we are above the
percolation threshold, our random lattice forms a giant cluster
within the hypercube and the typical length of loops is in the
number of sites $N=2^d$ of order $\log N$, where $N$ is the total
number of sites $N=2^d$, see \cite{bollobas_85}. Thus, the Bethe
lattice may be a good model for the topology of our sparse graph.

The main question addressed in our work will be the following:
Selective advantage in biological evolution is usually attributed
to higher reproductive success. If advantage due to high
individual fitness exceeds a certain threshold, a quasispecies is
formed around the site. This is the usual error-threshold
phenomenon. Here we ask, whether some factors related purely to
the structure or topology of the genome space may lead to similar
selective advantage, and if a certain threshold can be found
separating the adaptive and wandering regime of biological evolution.

\section{Model}

Recently, \cite{kol_sla_02} we have modelled evolution in a holey
landscape using the regular Bethe lattice. Now we will try to
represent it in a more precise way and take into account the
irregularity of the lattice. For each site we select whether the links
that leave  it lead  to another  `hospitable' site  of the
hypercube, or not. In the latter case, there  are some links
leading from the site to the `lethal' sites of the hypercube and
so all mutants that took this direction are doomed.

We start with a regular Bethe lattice with a connectivity $k$. We
construct the irregular Bethe lattice by randomly  assigning
lethal sites and removing all sites which are connected to the
rest of the lattice only through a lethal site. The evolution
process amounts to  diffusion on this lattice. Hops from a
viable site to any of its neighbors occur with an equal rate
$\mu$. Hops from lethal sites are prohibited. So, the edges to
lethal sites are ``dead ends'' or ``dangling bonds'', in the
language of condensed matter physics.

Let $i$ be a viable site. Let us denote $(i)$ as the set of viable
neighbors of $i$. $\kappa_i=|(i)|\leq k$ is the number of those
viable neighbors. For $k$ we suppose only that it is large enough
 to fulfill the previous inequality. The arbitrariness of the
choice of $k$ will be clarified later on. Intuitively, during
evolution the probability will flow out from $i$ to all its $k$
neighbors, but flow in only from $\kappa_i$ viable ones.
Therefore, we expect that a site with larger $\kappa_i$ is more
likely to gather individuals and form a quasispecies cloud. The
scope of this paper is to elaborate this intuitive picture in a
more formal manner.

If we denote $p_i(t)$ as a (relative) population\footnote{%
By ``population'' we mean  the infinite population
limit, the probability of finding a given individual in the site $i$ of
the lattice. We prefer the term ``relative'' population because of the
perturbations that will be added to the lattice and will destroy
the conservation of the overall relative population.}
of site $i$ at time $t$, we can write the following master equation
\begin{equation}\label{zaklad}
\dot{p}_i(t)=\sum_{j}T_{ij}\,p_j(t)
\end{equation}
where the matrix $T$ contains the effects of mutations and reproduction:
\begin{equation}
T_{ij}=\left\{\quad
\begin{array}{rl}
-\mu\,k+\zeta\quad&i=j\\
\mu\quad&j\in(i)\\
0\quad&\text{elsewhere.}
\end{array}
\right.
\end{equation}
The constant $\zeta$ is introduced artificially in order to keep
the total population constant. In fact, there are two causes of
the net population  outflow  which must be countered by the
$\zeta$ term. The first one is the presence of traps---lethal
sites that absorb the individuals. The second one comes from the
very topology of the Bethe lattice. Indeed, any finite Bethe
lattice has the rather counter-intuitive property that the number of
the surface sites is comparable to the number of bulk sites and
this property holds even in the limit of an infinite number of sites.
Therefore, there is always a net flow of probability toward the
surface. However, we are interested in the properties of the sites
deep in the bulk and the outflow toward the surface should be
considered as an artifact of the Bethe lattice approximation. To
sum up, we will fix the value of $\zeta$ later on in the
calculations, in order that the population remains fixed.

The dynamical matrix $T$ is in the model symmetric, and this enables
us to use the method of the resolvent in our calculations. The
symmetry results from the assumption of equal mutation rates,
$\mu$. This widely used approximation simplifies all the following
calculations and, since we are interested only in a stationary
state of the system, it is not considered to change the main
results. One can introduce the edge-dependent rates $\mu_{ij}$ and
then $\zeta$ would become a function of all $p_i(t)$ and
Eq.~(\ref{zaklad}) would generally become non-linear one. But this
approach goes beyond the scope of our article.

\section{Partitioning}

As in the previous paper \cite{kol_sla_02} we investigate the
formation of a localized state, now interpreted as a quasispecies
cloud, through the properties of the resolvent of the matrix $T$
\begin{equation}
\mathcal{G}(z)=(z-T)^{-1}\; .
\end{equation}

The idea of the calculation is quite simple: In the long-time
limit only the largest eigenvalue of the matrix $T$ survives, and
the corresponding eigenvector describes the stationary state of
the evolutionary system.

\begin{figure}[ht]
\begin{center}
\includegraphics[width=0.3\textwidth,height=0.3\textwidth]{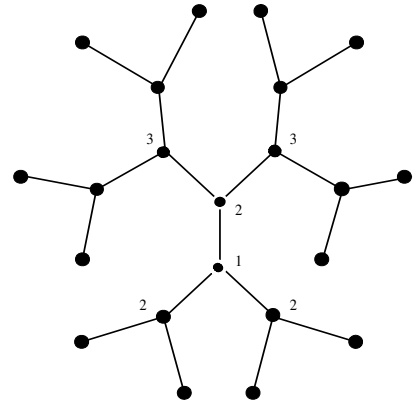}
\end{center}
\caption{Division of the Bethe lattice during the partitioning.}
\label{fig:bethepartitioning}
\end{figure}

In order to find the largest eigenvalue of $T$, we search for the
poles of an element of the resolvent matrix $\mathcal{G}$. These
are exactly the eigenvalues of $T$, no matter which element
of $\mathcal{G}$ we choose. And since we want to observe the
formation of the localized state around a specific site $i=1$, we
need the diagonal matrix element of the resolvent
\begin{equation}
G(z)=[\mathcal{G}(z)]_{11}
\end{equation}
which can be calculated using the partitioning (projector)
met\-hod \cite{lowdin_62}, explained in more detail in
\cite{kol_sla_02}. The loop-less structure of the Bethe lattice
greatly simplifies the treatment. We proceed essentially in two
steps, which are illustrated in the
Fig.~\ref{fig:bethepartitioning}. In the first step, we project
out of the site $i=1$ itself. The rest of the Bethe lattice splits
into disconnected branches. We find
\begin{equation}
G(z)=\frac{1}{z + \mu k -\zeta - \mu^{2}\sum_{j\in(1)}\Gamma_j(z)},
\label{eq:bg}
\end{equation}
where $\Gamma_j(z)$ is the diagonal element of the projected
resolvent on the terminal site of the branch starting with site
$j$. (The terminal sites are denoted as $2$ in
Fig.~\ref{fig:bethepartitioning}.)

In the second step, we calculate $\Gamma_j(z)$ by projecting out
the site $j$. We find, similar to the previous case,
\begin{equation}
\Gamma_j(z)=\frac{1}{z + \mu k -\zeta - \mu^{2}\sum_{l\in(j)\backslash\{1\}}\Gamma_{l}(z)}\; .
\label{eq:forgamma}
\end{equation}
Some of the sites $l\in(j)\backslash\{1\}$ are denoted by $3$ in
 Fig.~\ref{fig:bethepartitioning}. Iterating the equation
(\ref{eq:forgamma}) we could in principle calculate the resolvents
on the terminal sites and inserting them into Eq.~(\ref{eq:bg}) 
obtain the desired quantity.

This procedure works well in an infinite regular Bethe lattice, where
$\Gamma_j(z)$ for $j$ deep in the bulk does not depend on the site
index $j$ and (\ref{eq:forgamma}) represents in fact a closed
equation for $\Gamma(z)\equiv \Gamma_j(z)\;\forall j$. However,
this procedure cannot be directly applied in the case of an irregular
Bethe lattice. Nevertheless, it is a good starting point for an
approximation we will describe in the following.

\section{Effective medium approximation}

The mean-field-type treatment of disordered solids was developed a
long time ago within the coherent potential approximation (CPA)
\cite{vel_kir_ehr_68}. In the theory of sparse random matrices it
was elaborated using the replica method
\cite{bi_mo_99,monasson_99,sem_cug_02} and called the effective medium
approximation (EMA). In this section we will use the EMA for the
irregular Bethe lattice without using the replica trick.

The main idea relies on a simple observation that the sum
$\sum_{l\in(j)\backslash\{1\}}\Gamma_{l}(z)$ containing
$\kappa_j-1$ terms can be replaced by its average value for  large
$\kappa_j$, thus neglecting fluctuations. Indeed, it was
proved that the CPA is exact in the limit of infinite
connectivity \cite{jan_vol_92}. Therefore, our version of the EMA
amounts to approximating
\begin{equation}
\sum_{l\in(j)\backslash\{1\}}\Gamma_{l}(z)\simeq (\kappa_j-1)\,\langle
\Gamma(z)\rangle\; .
\label{eq:approximation}
\end{equation}
In order to close the equations, we must average expression
(\ref{eq:forgamma})
over the probability distribution of connectivities $P(\kappa)$
\begin{equation}
\langle \Gamma(z)\rangle =\sum_\kappa
\frac{P(\kappa)}{z + \mu k -\zeta -
\mu^{2}(\kappa-1)\langle\Gamma(z)\rangle}\; .
\label{eq:forgammaeveraged}
\end{equation}
The latter equation (\ref{eq:forgammaeveraged}) is the core of the
EMA. When we insert its solution $\langle\Gamma(z)\rangle$ into
(\ref{eq:bg}) and perform again the average over connectivities,
we obtain the averaged diagonal element of the resolvent $\langle
G(z)\rangle$, which is now site-independent. The parameter $\zeta$
represents the shift in the variable $z$ and should be adjusted so
that the upper edge of the support of the imaginary part of
$\langle G(z)\rangle$ (i.e. the density of states) lies at $z=0$.
This expresses the requirement of the conservation of the
population size.
\begin{figure}[ht]
\includegraphics[bb=80 470 370 680,width=0.45\textwidth]
{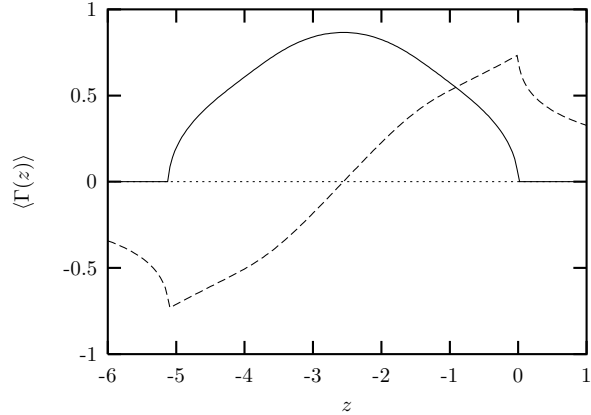} \caption{The averaged resolvent
at terminal site $\langle\Gamma(z)\rangle$ for the irregular Bethe
lattice with connectivity distribution defined in
(\ref{eq:kappadistribution}), where $p=\frac{1}{2}$. Full line:
imaginary part. Dashed line: real part. }
\label{fig:resolvent-ema}
\end{figure}

As an illustration we show in Fig.~\ref{fig:resolvent-ema} the real and
imaginary parts of $\langle\Gamma(z)\rangle$ for the connectivity
distribution chosen as
\begin{equation}
P(\kappa)=p\,\delta(\kappa-2)+(1-p)\,\delta(\kappa-3)
\label{eq:kappadistribution}
\end{equation}
for $0\le p\le 1$. We can see that the imaginary part of
$\langle\Gamma(z)\rangle$ approaches zero as $z^{1/2}$ at the band
edge and the real part approaches a finite limit. From the
technical point of view, the finiteness of the limit is the source
of the transition between localized and delocalized states.

\section{Single defect}

\begin{figure}[ht]
 \begin{center}
 \includegraphics[width=0.3\textwidth]{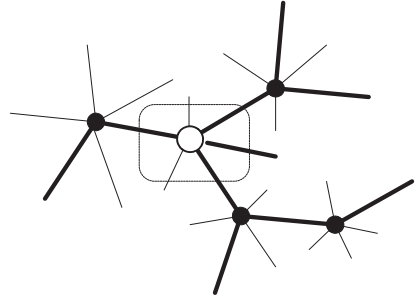}
 \end{center}
 \caption{The ``defect''
  site  with a different  connectivity. Thin links represent
lethal mutations, the thick ones mutations to viable sites.}
\label{fig:fpgp}
\end{figure}

So far we have investigated the Bethe lattice as an averaged
homogeneous effective medium. Now we investigate the behavior of
the resolvent at a site $i=1$ with a specific connectivity
$\kappa_1$. The situation is schematically depicted in
Fig.~\ref{fig:fpgp}.

This approach has a biological motivation. What we see in nature
is that there are frequent groups of closely related  animals.
These groups form taxonomic classes or, better said, the classes
are defined as the groups of such closely related individuals. In
the following, we examine if the clustering of individuals, modelled
as the addition of the site with largest connectivity, leads to some
observable chan\-ges in the quasispecies evolutionary process.

\begin{figure}[ht]
 \includegraphics[bb=86 472 370 680,width=0.45\textwidth]{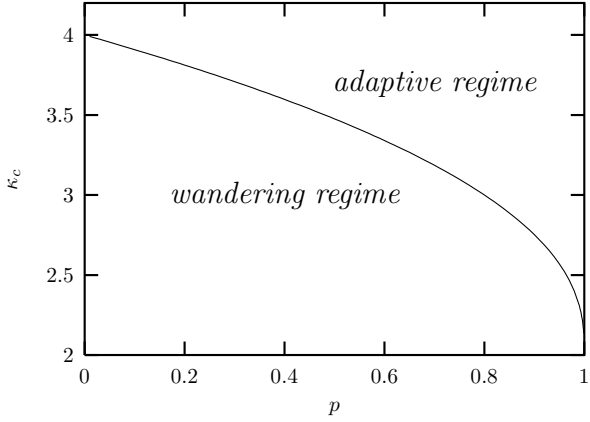}
 \caption{  The
 dependence of  the critical value of  the connectivity, $\kappa_{c}$,
 on $p$, for connectivity distribution (\ref{eq:kappadistribution}).}
\label{fig:kappac}
\end{figure}

The on-site resolvent corresponding to this site can be found from
Eq.~(\ref{eq:bg}), where the resolvents at the terminal sites are
approximated as in (\ref{eq:approximation}). This is the essence
of the single-defect approximation (SDA). Hence
\begin{equation}
G_{\rm SDA}(z)=\frac{1}{z + \mu k -\zeta -
\mu^{2}\kappa_1\langle\Gamma(z)\rangle}\; .
\label{eq:singledefect}
\end{equation}
A state localized at this site exists, if the resolvent $G_{\rm
SDA}(z)$ has a pole for a real positive $z$. This is equivalent to
the condition
\begin{equation}
\kappa_1>\kappa_{c}\equiv\frac{\mu k-\zeta}{\mu^2}\,
\frac{1}{{\rm Re}\langle\Gamma(0)\rangle} \; .
\label{eq:kappalocalized}
\end{equation}
Therefore, the state is localized and the quasispecies cloud is
formed for integer $\kappa_1$ greater than $\kappa_c$. This
corresponds to the adaptive regime of the evolutionary  process.
We use as an example again the distribution
(\ref{eq:kappadistribution}) and show in the Fig.~\ref{fig:kappac}
the dependence of the threshold  $\kappa_c$ on the parameter $p$
of the probability distribution.

We will consider yet another type of ``defect'' in the
effective medium. Imagine that the site investigated, $i=1$,
corresponds to a genome of length different to the average.
This difference is translated in our model as a choice of
different total connectivity $k$ in the site $i$ (including edges
to both viable and lethal sites). So, we may replace $k\to
k+\lambda$ in Eq.~(\ref{eq:singledefect}). Then we may investigate
the transition from the wandering to the adaptive regime when varying two
parameters $\lambda$ and $\kappa_1$. We will see that the
individuals added due to the $\zeta$ term in Eq.~(\ref{zaklad})
are redistributed in the lattice in such a manner that the total
population changes. This is only due to addition of the ``defect''
and the localization of individuals in its vicinity. The condition for
the existence of a localized state, i.e. for the adaptive regime,
is analogous to (\ref{eq:kappalocalized}) and can be written as
\begin{equation}
\lambda<\lambda_c\equiv\mu\kappa_1\,
{\rm Re}\langle\Gamma(0)\rangle -\frac{\mu k -\zeta}{\mu}\; .
\label{eq:lambdalocalized}
\end{equation}
\begin{figure}[ht]
 \includegraphics[bb=86 472 370 680,width=0.45\textwidth]
{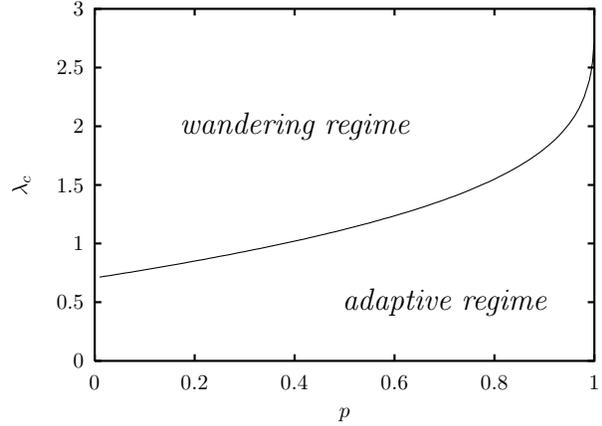}
\caption{
The dependence of the  critical value $\lambda_{c}$ on the probability
$p$, for distribution (\ref{eq:kappadistribution}) and $\kappa_1=5$.}
\label{fig:lambdac}
\end{figure}

We show in Fig.~\ref{fig:lambdac} an example of the dependence of
$\lambda_c$ for the same connectivity distribution
(\ref{eq:kappadistribution}) as in previous cases, and for a
specific, fixed choice of $\kappa_1$.

\section{Conclusions}

We modelled the complex fitness landscape of biological evolution
with an irregular Bethe lattice. The formation of localized
quasispecies in the adaptive regime was observed via the occurrence of the
isolated pole in the on-site resolvent. Using the effective medium
approximation we calculated the disorder-averaged resolvent and
within the single-defect approximation we investigated the
localization.

We found that the quasispecies is formed if the site is connected
to a sufficiently large number of viable sites. We found the
condition for the critical value of the connectivity $\kappa_c$
separating the adaptive evolutionary regime for $\kappa>\kappa_c$
and the wandering regime for $\kappa<\kappa_c$. As expected,
$\kappa_c$ grows with the average connectivity and qualitatively
speaking the quasispecies are formed at such sites that have
sufficiently larger number of viable neighbors than average. We
may interpret it as a purely topological selective advantage: The
species has better chance to survive not because of its individual
reproductive abilities, but because it is less vulnerable to
random alterations of the genetic code.

We investigated also another deviation from the mean, connected to
the overall number of neighbors. In more biological language it
corresponds to the length of the genetic code. We observed that
the adaptive regime is favored for lower values of the
connectivity, i.e. for shorter genomes, if the number
of viable neighbour sites is considered constant. This type of selective
advantage is therefore also of topological origin and has a similar
biological interpretation to that presented above. Indeed, longer genome
means higher probability of a lethal mutation.

In both cases it is important to note that it is the {\it
deviation} from the average topology which makes the selective
advantage work and which leads to the formation of localized
states. So, it is the (sufficiently strong) topological disorder
that is responsible for the formation of quasispecies.

We may conclude by summarizing that without resort to individual
reproductive capacities, biological evolution favors genomes which
are shorter and more robust to random mutations. This has one more
important implication; a genome, which can be easily mutated
without affecting the death of its carrier, means also a
less-defined species. One may therefore predict that successful
species will exist in a broad variety of slightly different
sub-species. This effect is caused by the selective advantage of
certain topologies of the genome space. Observation of the
variability within a single species may therefore say something of
the relative importance of topological selective advantage in comparison
 to individual reproductive success.

\begin{acknowledgement}
The authors would like  to thank to Jan Ma\v{s}ek  for critical
reading of the manuscript. Michal Kol\'{a}\v{r} would like to
thank to Vladislav \v{C}\'{a}pek  for fruitful  discussions and
to Anton  Marko\v{s} for helpful  discussions on the  topic of
biological evolution,  too.
\end{acknowledgement}

\end{document}